\documentstyle{caosp}
\begin{document}
\firstpage{446}
\hauthor{S.M. Andrievsky {\it et al.}}
\title{Accurate LTE abundances for some $\lambda$ Boo stars}
\author{S.M. Andrievsky \inst{1} \and I.V. Chernyshova \inst{1}
\and V.G. Klochkova  \inst{2} \and V.E. Panchuk \inst{2}}
\institute{Department of Astronomy, Odessa State University, Shevchenko Park,
270014, Odessa, Ukraine\\
\and
Special Astronomical Observatory, Russian Academy of Sciences,
Nizhny Arkhyz, Russia}
\date{\today}
\maketitle
\begin{abstract}
High-resolution and high S/N CCD spectra were analyzed to determine
accurate LTE abundances in four $\lambda$ Boo stars: $\pi^{1}$ Ori, 29 Cyg,
HR 8203 and 15 And. In general, 14 chemical elements were investigated. The main
results are the following: all stars have a strong deficiency of the majority
of investigated metals. Oxygen exhibits a moderate deficiency.
The carbon abundance is close to the solar one.

The results obtained support an accretion/diffusion model, which is currently
adopted for the explanation of the $\lambda$ Boo phenomenon.
\end{abstract}
\section{Introduction}
Among the unresolved problems of stellar astrophysics, there is one
linked with the $\lambda$ Boo phenomenon.
A comprehensive review of the $\lambda$ Boo phenomenon was recently provided
by St\"{u}renburg (1993) and by Paunzen et al. (1997).

To explain the $\lambda$ Boo phenomenon, Venn and Lambert (1990) adopted
an accretion hypothesis. According to that hypothesis, the chemical peculiarity
of $\lambda$ Boo stars originates due to the presence of a circumstellar shell.
The circumstellar shell consists of two phases: gas and dust grains.
The dust grains accumulate metals having a high condensation temperature
(e.g. Si, Fe), but elements with lower condensation temperature (C, N)
remain in the gaseous phase. Depleted gas from the circumstellar envelope
is accreted by the star, while dust grains drift out of the shell due to 
radiative pressure.

Further studies of the proposed accretion scenario were made by
Charbonneau (1991, 1993), who combined it with the theory of diffusion.
Attempts to derive accurate elemental abundances in the atmospheres of
$\lambda$ Boo stars were undertaken in several works (Venn and Lambert, 1990;
St\"{u}renburg, 1993, etc).
\section{Observation}
The CCD spectra have been obtained with the \'echelle spectrometer LYNX on
the 6m telescope (Special Astrophysical Observatory of the Russian Academy
of Sciences, Russia, Northern Caucasus). The detailed description of the
spectrometer is given by Panchuk et al. (1993). The resolving  power was
24000, S/N $\approx$ 100.
\section{Atmospheric parameters}
Temperatures and gravities of the programme stars were estimated using the
(b-y)-c$_{1}$ grid by Kurucz (1991). Str\"omgren colours were selected from
Hauck \& Mermilliod (1985), $v\sin i$ values are from the catalogue of
Paunzen et al. (1997). For all stars a microturbulent
velocity of 3 km\,s$^{-1}$ was adopted. The results are given in Table 2.
\begin{table}[t1]
\small
\begin{center}
\caption{Characteristics of programme stars.}
\label{t2}
\begin{tabular}{lccccc}
\hline
 Star    &(b-y)  & c$_{1}$&$v\sin i$, km\,s$^{-1}$&T$_{\rm eff}$,K&$\log g$ \\
\hline
 HR1570  &0.044  &1.007   & 105                &  8750     &  4.2   \\
 HR7736  &0.101  &0.927   & 80                 &  8000     &  4.2   \\
 HR8203  &0.093  &0.940   & 65                 &  8300     &  4.2   \\
 HR8947  &0.056  &1.072   & 100                &  9000     &  4.1   \\
\hline
\end{tabular}
\end{center}
\end{table}
\section{Method}
To derive elemental abundances, we applied the spectral-synthesis technique 
(STARSP code by Tsymbal, 1996; atmosphere models come from Kurucz, 1992). 
Oscillator strengths for the investigated lines and blends were corrected
by compariing the solar synthetic spectrum (solar model from Kurucz's grid,
V$_{t}$=1 km\,s$^{-1}$ and solar abundances from Grevesse \& Noels, 1993) with
the solar flux spectrum (Kurucz et al. 1984).
\section{Results}
In Table 2 we give the individual abundances of the programme stars in the form
[El/H].
\begin{table}[t2]
\small
\begin{center}
\caption{Abundances of the programme stars.}
\label{t3}
\begin{tabular}{lcccc}
\hline
El. &$\pi^{1}$ Ori&29 Cyg & HR8203& 15 And\\
\hline
C   &-0.2         &  -0.1 & -0.2  &  0.0 \\
O   &-0.5         &  -0.5 & -0.5  & -0.3 \\
Na  &-0.5         &  -0.4 & -0.5  &      \\
Mg  &-0.8         &  -1.5 & -1.0  & -0.5 \\
Si  &-1.3         &  -1.0 & -0.8  & -1.0 \\
S   &-0.4         &       & +0.4  &      \\
Ca  &-1.0         &  -0.8 & -0.8  & -0.6 \\
Sc  &-0.8         &  -0.5 & -0.9  & -0.6 \\
Ti  &-0.7         &  -1.0 & -0.8  & -0.6 \\
Cr  &-0.6         &  -1.2 & -0.5  & -0.4 \\
Fe  &-1.0         &  -1.3 & -0.7  & -0.8 \\
Ni  &-0.5         &  -0.6 & -0.4  &      \\
Ba  &-0.5         &  -0.4 &  0.0  &      \\
\hline
\end{tabular}
\end{center}
\end{table}
In the present study we confidently confirmed that carbon has an
approximately normal abundance in all investigated stars. Oxygen is moderately 
deficient. There is also no doubt that most of the heavier elements are strongly
deficient in the atmospheres of $\lambda$ Boo stars.

As it was realized by Venn and Lambert (1990), relative abundances must depend
upon the condensation temperature.
Our result strongly supports the supposition that the gas from the
circumstellar shell, enriched in the elements with low $T_{\rm cond}$ values (C,
O, S), is preferentially accreted by the star, while elements with a high
condensation temperature (Ca, Ti, Fe, etc) and locked in dust grains,
do not reach the stellar atmosphere.

\end{document}